\documentclass[aps,prl,twocolumn,nofootinbib,showpacs,longbibliography]{revtex4-2}
\usepackage{bm,amsmath,amssymb,graphicx}
\usepackage[colorlinks,linkcolor=blue,urlcolor=blue]{hyperref}
\newcommand{\iR}{{\rm i}}

\begin{document}

\title{Elementary particles with nonzero spin must be massless}

\author{Hans Christian \"Ottinger}
\email[]{hco@mat.ethz.ch}
\homepage[]{www.polyphys.mat.ethz.ch}
\affiliation{ETH Z\"urich, Quantum Center and Department of Materials, HCP F 43.1, CH-8093 Z\"urich, Switzerland}

\date{\today}

\begin{abstract}
We present an ontological argument why elementary particles with nonzero spin must be massless. This argument implies that, from an ontological perspective, the massive quarks and leptons of the standard model cannot be elementary particles. This conclusion is less disquieting than it might seem at first sight because the Higgs mechanism entails that not only the masses of the W and Z bosons but also the masses of quarks and leptons arise from the interaction of massless elementary particles with the vacuum expectation value of the Higgs field, which is a result of symmetry breaking.
\end{abstract}

\maketitle

\section{Introduction}
We here use the term \emph{elementary particle} in two different shades: pragmatic and ontological. From a \emph{pragmatic perspective}, elementary particles are what physicists use to describe and analyze experimental observations at particle colliders. More explicitly, these are the elementary particles currently figuring in the standard model: quarks and leptons as the fermionic constituents of matter as well as photons, the W and Z bosons and gluons as gauge vector bosons associated with electroweak and strong interactions and, finally, the Higgs boson. Less well-established are gravitons; there are appealing reasons to believe that these bosons mediating gravitational interactions should be gauge vector bosons of the same kind as those used so successfully for describing all other interactions \cite{Utiyama56,Yang74,Camenzind77a,Camenzind77b,Camenzind78,IvanenkoSar83,CapozzielloDeLau11,hco231,hco240,hco252}.

From an \emph{ontological perspective}, elementary particles are what ultimately exists in nature, what all stuff in the universe is made from. A key assumption is that elementary particles in the ontological sense must be entirely characterized by their intrinsic properties.

Fundamental particles of neither shade are particles in a classical sense. The theory of elementary particles is known as quantum field theory, bringing up the famous question ``particles or fields?'' Among philosophers of physics, a field interpretation seems to be prevailing, presumably as a result of serious no-go theorems for particles, which mostly establish the non-localizability of quantum particles \cite{Malament96ip,Hegerfeldt98ip,Hegerfeldt98,HalvorsonClifton02,Oldofredi18}.

Of course, arguments against particles are not automatically arguments in favor of fields because the question ``particles or fields?'' may actually be misleading. These might neither be all options, nor must they be mutually exclusive. The term \emph{field quanta} indicates how a bridge between particles and fields may be built. Our task is to find positive arguments for suitably refined quantum versions of particles or fields that help us to think and talk about experimental observations, both in a pragmatic and in an ontological sense. Philosophers are well advised to be pragmatic and stay in contact with the impressively successful standard model of particle physics \cite{Wallace06}. Conversely, physicists are well advised to make use of conceptual clarifications provided by philosophy \cite{hcoqft}. The purpose of the present note is to bring together physical and philosophical arguments and concepts to advance our understanding of elementary particles.

Wigner's \cite{Wigner39} classification of the irreducible representations of the Poincar{\'e} group can be used to argue that all relativistic particles possess a nonnegative mass $m$ and an integer or half-integer spin $s$. The purpose of this work is to present and discuss an ontological argument why, for elementary particles, at least one of these two properties, $m$ or $s$, must be zero. In the standard model, the only particle with $s=0$ is the Higgs boson. All other elementary particles possess $s=1/2$ (quarks and leptons) or $s=1$ (gauge vector bosons) and hence must be massless. They acquire mass only by interactions, in particular, with the vacuum expectation value of the Higgs field. Therefore, symmetry breaking is essential for mass to arise, both for the W and Z bosons and for the quarks and leptons of the standard model.

\section{Fock space}
The concept of Fock space \cite{Fock32} realizes the idea of discrete entities that may be aggregated \cite{Teller}. One may think of these entities as quanta, in particular as field quanta, or as indistinguishable quantum particles. An underlying tensor product structure implies that the aggregation does not account for any interaction effects between the entities. It is hence natural to think of the entities associated with Fock space as free quantum particles, which are naturally taken to be in momentum eigenstates. If we assume that the field quanta are contained in a large but finite volume then the set of possible momentum vectors is discrete.  Note that Fock space comes with a natural basis: every base vector is associated with a well-defined content of free elementary particles in momentum eigenstates. Interactions may be introduced as collisions between free particles \cite{hcoqft}.

According to Heisenberg's uncertainty relation, sharp momentum values for free quantum particles imply that there is no spatial resolution. There is no way to predict the position of a free quantum particle in a momentum eigenstate. However, according to the standard model, the interactions between elementary particles may be considered as strictly local collisions involving either three or four particles at a particular point in space and time \cite{hcoqft}. In a typical collider experiment, a high-energy collision is followed by a sequence of correlated low-energy collisions, and certain particles arising in the latter collisions are detected (through further collision events). This is how the typical particle tracks emerging from the point of a high-energy collision arise in collider experiments \cite{hco243}. We cannot predict the position of a free particle at a particular time, but particles reveal their positions through a sequence of collisions occurring along discrete trajectories.

In order to formulate a quantum field theory we need (i) a list of elementary particles and (ii) the collision rules for these particles. The list of elementary free particles determines the underlying Fock space, the collision rules express the interaction Hamiltonian, which the Fock space knows nothing about. The free particles of the Fock space constitute the elementary particles in the ontological sense, provided that they are characterized by intrinsic properties. The elementary particles in the pragmatic sense result from the inevitable pervading interactions between free particles.

Quantum field theory is usually formulated as a Hamiltonian theory, that is, as a system with reversible dynamics. Physicists quite generally seem to make the implicit metaphysical assumption that the most fundamental theory of nature must be reversible. One can conversely argue that, as nobody expects the standard model to be a most fundamental theory of nature, it should better be formulated in terms of dissipative equations. Indeed, dissipative quantum field theory (DQFT) provides a robust formulation of particle physics, where regularization at unresolved small scales is provided by diffusive smearing \cite{hcoqft}. Of course, then a detailed dissipation mechanism predominantly active on small length scales needs to be formulated in addition to the Fock space and the interaction Hamiltonian.

In DQFT, physical states are represented by density matrices on a Fock space evolving according to quantum master equations \cite{BreuerPetru,Weiss}. The distinction between a free theory and interactions gains deep physical meaning because (i) only the free particles are subject to dissipation and (ii) interactions are interpreted as discrete collision events between continuously evolving free particles. The free vacuum state, which is the Fock state not containing any free particles, is conceptually different from any physical state of the interacting system, represented by a density matrix and accounting for both quantum and thermal fluctuations. In particular, the free (or ontological) vacuum state is conceptually different from the physical (or pragmatic) vacuum state, in which free particles are constantly created and annihilated in interactions or collisions among them.

DQFT is more robust than purely reversible formulations of quantum field theory, and it eliminates the inconsistency of the interaction picture associated with Haag's theorem \cite{Ruetsche,Duncan}. By unraveling the quantum master equation of DQFT in terms of two stochastic processes, one can obtain an appealing ontological interpretation \cite{hcoqft,hco243,hcox1} and a new simulation technique for elementary particle physics \cite{hcoqft,hco248,hco251} that provides an alternative to the usual Monte Carlo simulations of lattice gauge theories \cite{Wilson74,DuaneKogut86,Gottliebetal87,Gockeleretal98,Bazavovetal10}.

\section{The ontological argument}
In order to set up the Fock space of elementary particles in the ontological sense we need to characterize the free particles by their intrinsic properties. The simplest case is that of scalar free particles with spin $s=0$ and mass $m$, which can be fully characterized by their momenta. The energy of the free particles is given by the relativistic energy-momentum relation. Natural Fock space basis vectors can be constructed by acting with the creation operators $a_{\bm k}^\dag$ on the free vacuum state $\left| 0 \right\rangle$, where ${\bm k}$ refers to the possible momenta of the free particles created by these operators. Particles with spin $s=1/2$, say electrons, require creation operators $b_{\bm k}^{\sigma \, \dag}$ with the additional label $\sigma$ taking values $\pm 1/2$ for specifying the spin state (and, according to Dirac's theory, additional creation operators for their antiparticles). This characterization of different spin components requires a particular direction in space, typically called the $3$ or $z$ direction. For ontological reasons, we insist that an elementary particle should be characterized by intrinsic properties without making any reference to an arbitrarily chosen direction in space.

One might choose to be satisfied if the final results of a theory do not depend on the particular choice of a reference direction in space. In this note, I suggest to insist that it must be possible to characterize elementary particles by intrinsic properties that are independent of any choice of an arbitrary direction. If elementary particles are what ultimately exists in nature, they must be fully characterized by their intrinsic properties rather than conventions chosen by an observer. This is the fundamental ontological assumption of the present note.

If we want to describe particles with spin in the usual way then we need a preferred direction in space associated with a particle. The basic idea for avoiding arbitrariness is that a distinguished direction in space is given by the direction of motion of the particle. We must make sure that an elementary particle with nonzero spin must always be in an intrinsic state of motion. Only massless particles cannot be at rest in any frame of reference and hence always come with an intrinsic direction of motion.

Massless particles always move with the speed of light, $c$. Their direction of motion is not affected by a change the frame of reference, provided that this change is characterized by a Lorentz transformation with a velocity vector ${\bm v}$ with $|{\bm v}|<c$. In particular, the direction of motion of a massless particle cannot be reversed by changing the frame of reference. As is known from the Doppler effect for light, the only effect of changing the frame of reference is given by a change of the units of energy by the Lorentz factor. In the following, we assume a fixed Minkowski system with the corresponding units of energy.

For the further discussion, we distinguish three types of quantum particles: f-particles, o-particles, and p-particles. The f-particles are the free particles associated with the creation operators of a Fock space. If f-particles are fully characterized by their intrinsic properties, they qualify as o-particles or ontological particles. The p-particles, or physical particles in the pragmatic sense, are the ones used for analyzing collider experiments; these particles incorporate the inevitable presence of interaction effects.

\section{Essentials of electroweak theory}
For further clarification, we consider the Glashow-Weinberg-Salam theory of electroweak interactions and the Englert–Brout–Higgs–Guralnik–Hagen–Kibble mechanism (or, briefly, Higgs mechanism) for generating mass. Our discussion of the essentials is based on Sections 20.1 and 20.2 of \cite{PeskinSchroeder} and Section 21.3 of \cite{WeinbergQFT2}, which may be consulted for further details. As the masses of gauge bosons, quarks and leptons all arise from interactions, we begin our discussion with the massless free theories.

\subsection{Massless free leptons}
Second quantization of the Dirac equation for the electron is usually associated with four creation operators for the spin states of the electron ($b$) and the positron ($d$) \cite{hcoqft},
\begin{equation}\label{freeelectronpositron}
    b^{\sigma \, \dag}_{\bm{k}} \,, \quad d^{\sigma \, \dag}_{\bm{k}} \,,
\end{equation}
where $\bm{k}$ indicates momentum and $\sigma = \pm 1/2$ represents the spin component along a fixed direction in space. The creation operators (\ref{freeelectronpositron}) define massless f-particles. To arrive at o-particles, we consider the spin component along the momentum direction, which is known as helicity. For massless free fermions, the helicity is a well-defined intrinsic property, which is a Lorentz invariant constant of motion. The o-particles are given by the creation operators for left- and right-handed electrons and positrons,
\begin{equation}\label{oelectronpositron}
    b^{{\rm L} \, \dag}_{\bm{k}} \,, \quad b^{{\rm R} \, \dag}_{\bm{k}} \,, \quad
    d^{{\rm L} \, \dag}_{\bm{k}} \,, \quad d^{{\rm R} \, \dag}_{\bm{k}} \,.
\end{equation}
For massless particles, helicity coincides with chirality. More generally, projections on chiral states are defined such that they are Lorentz invariant even for massive particles.

Although inclusion of all lepton and quark generations is essential for eliminating gauge anomalies from the electroweak theory (see, e.g., pp.\,705--707 of \cite{PeskinSchroeder}), we here consider only electrons and positrons because all other fermions can be treated in a similar way.

\subsection{Massless free gauge vector bosons}
The theory of electroweak interactions is a gauge theory based on the symmetry groups $SU(2)$ and $U(1)$ associated weak isospin and weak hypercharge, respectively, where the latter is a generalization of the electric charge. The corresponding creation operators are conveniently taken as
\begin{equation}\label{EWfreegaugebosons}
    w^{a \mu \, \dag}_{\bm{k}} \,, \quad y^{\mu \, \dag}_{\bm{k}} \,,
\end{equation}
where $\bm{k}$ indicates momentum, $\mu$ represents space-time components and $a$ labels the three generators of the symmetry group $SU(2)$, typically chosen as the Pauli matrices. The creation operators (\ref{EWfreegaugebosons}) define massless f-particles but, as they depend on our choice of coordinates, no o-particles.

We first consider the creation operators $y^{\mu \, \dag}_{\bm{k}}$ of the free vector bosons associated with weak hypercharge. As a massless particle moves with the speed of light, its momentum $\bm{k} \neq \bm{0}$ can be used to define an intrinsic $3$ direction. As there still is a rotational degree of freedom associated with the choice of the $1$ and $2$ directions, we pass from transverse to circular polarization states,
\begin{equation}\label{pmphotondef}
    y^{{\rm +} \, \dag}_{\bm{k}} = \frac{1}{\sqrt{2}} ( y^{1 \, \dag}_{\bm{k}} + \iR y^{2 \, \dag}_{\bm{k}} )  , \quad
    y^{{\rm -} \, \dag}_{\bm{k}} = \frac{1}{\sqrt{2}} ( y^{1 \, \dag}_{\bm{k}} - \iR y^{2 \, \dag}_{\bm{k}} ) .
\end{equation}
Except for an irrelevant phase factor, the creation operators (\ref{pmphotondef}) are defined unambiguously, independent of the choice of the $1$ and $2$ directions. In the same spirit, we can immunize longitudinal and temporal polarization states to Lorentz boosts along the momentum direction by passing to left (gauche) and right (droite) polarization states,
\begin{equation}\label{gdphotondef}
    y^{{\rm g} \, \dag}_{\bm{k}} = \frac{1}{\sqrt{2}} ( y^{0 \, \dag}_{\bm{k}} + y^{3 \, \dag}_{\bm{k}} ) , \quad
    y^{{\rm d} \, \dag}_{\bm{k}} = \frac{\iR}{\sqrt{2}} ( y^{0 \, \dag}_{\bm{k}} - y^{3 \, \dag}_{\bm{k}} ) .
\end{equation}
By passing from the space-time components of the creation operators $y^{\mu \, \dag}_{\bm{k}}$ to the polarization states $\{ +, -, {\rm g}, {\rm d} \}$ introduced in (\ref{pmphotondef}) and (\ref{gdphotondef}), we have obtained the creation operators for the o-particles mediating the interactions associated with weak hypercharge. These operators create vector bosons of the ontological type.

By similar strategies, all ambiguities can be removed from the creation operators $w^{a \mu \, \dag}_{\bm{k}}$ for the set of free gauge vector bosons associated with weak isospin. The passage from space-time components to polarization states works in exactly the same way as for $y^{\mu \, \dag}_{\bm{k}}$. As $a$ in the creation operators $w^{a \mu \, \dag}_{\bm{k}}$ labels the three Pauli matrices representing weak isospin, an unambiguous label $3$ is associated with the direction of symmetry breaking for the Higgs boson. The transverse isospin components $1$ and $2$ can then be treated as in (\ref{pmphotondef}).

\subsection{Massive physical gauge vector bosons}
At this point, we introduce the Higgs boson to generate the observed masses of the gauge vector bosons. It is not straightforward to include the Higgs boson into the theory because it is unstable. As a result, symmetry breaking occurs and a constant effective background field arises. The magnitude $v$ of this effective field has dimensions of mass and is a free parameter of the theory. The direction of the background field provides the previously mentioned distinguished direction in isospin space.

The existence of massless photons in the pragmatic sense poses the following challenge: How can we identify a vector boson that is not affected by the Higgs mechanism? For that purpose, we need to consider the following creation operators constructed by mixing the operators $y^{\mu \, \dag}_{\bm{k}}$ and $w^{3 \mu \, \dag}_{\bm{k}}$, which create particles with the same quantum numbers,
\begin{eqnarray}
   a^{\mu \, \dag}_{\bm{k}} &=&
   \cos\theta_w \, y^{\mu \, \dag}_{\bm{k}} - \sin\theta_w \, w^{3 \mu \, \dag}_{\bm{k}} , \\
   z^{\mu \, \dag}_{\bm{k}} &=&
   \sin\theta_w \, y^{\mu \, \dag}_{\bm{k}} + \cos\theta_w \, w^{3 \mu \, \dag}_{\bm{k}} ,
\end{eqnarray}
where $\theta_w$ is the weak mixing angle. For properly chosen $\theta_w$, the vector bosons created by $a^{\mu \, \dag}_{\bm{k}}$ are not affected by the Higgs mechanism and should hence be considered as photons. The operators $z^{\mu \, \dag}_{\bm{k}}$ create $Z$ bosons. The proper $\theta_w$ is determined by the ratio of the coupling constants $g$ and $g'$ of the gauge theories with the symmetry groups $SU(2)$ and $U(1)$ associated weak isospin and weak hypercharge,
\begin{equation}\label{wmixangle}
   \cos\theta_w = \frac{g}{\sqrt{g^2+{g'}^2}} , \qquad
   \sin\theta_w = \frac{g'}{\sqrt{g^2+{g'}^2}} .
\end{equation}
A detailed elaboration of the Higgs mechanism leads to the following masses of the $W$ and $Z$ bosons \cite{PeskinSchroeder,WeinbergQFT2},
\begin{equation}\label{WZmasses}
   m_W = \frac{1}{2} g v , \qquad m_Z = \frac{m_W}{\cos\theta_w} > m_W .
\end{equation}
Experimental results suggest $\theta_w \approx 29$\textdegree.

\subsection{Massive physical leptons}
The masses of leptons and quarks cannot be implemented by simple mass terms in the Hamiltonian or Lagrangian because such terms would violate gauge invariance. This invariance can be achieved by a suitable interaction of the fermions with the Higgs field. For example, the resulting mass of the electron is given by (see (20.100) of \cite{PeskinSchroeder} or (21.3.31) of \cite{WeinbergQFT2})
\begin{equation}\label{emass}
   m_e = g_e v ,
\end{equation}
where $g_e$ is the dimensionless coupling constant of the electron-Higgs interaction. As each fermion has its own coupling constant, the masses of all leptons and quarks are model parameters. The complete list of free parameters of the electroweak theory may be taken as the masses of the Higgs particle, the $W$ and $Z$ bosons and all leptons and quarks plus one coupling constant, which may conveniently be chosen as the strength of the well-known electromagnetic interactions.

\section{Summary and conclusions}
From 1920 on, the proton, the neutron and the electron have been considered as elementary particles for almost half a century. In 1964 Gell-Mann \cite{GellMann64} proposed that protons and neutrons consist of three quarks and, five years later, inelastic scattering experiments revealed that protons and neutrons indeed are composed of pointlike constituents \cite{BjorkenPaschos69}. The nucleons had lost there status as truly elementary particles, but not as elementary particles in a pragmatic sense. For example, even today particle physicists would not hesitate to say that the Large Hadron Collider at CERN is primarily used as a proton collider or to assign particle tracks in a bubble chamber to protons.

In this note we have argued that Gell-Mann's quarks as massive fermions cannot be elementary particles in an ontological sense either. More generally, an elementary particle with nonzero spin must have zero mass so that its direction of motion implies a well-defined direction in space, along which its intrinsic helicity is defined. We have elaborated the details of our ontological argument in terms of free, ontological, and pragmatic particles in the Fock space setting of quantum field theory. All pragmatic or physical particles with nonzero spin acquire mass through interactions with the Higgs boson. There is no reason to expect that elementary particles of the ontological type are in eigenstates of interaction-generated mass.

Textbooks on quantum field theory focus on the questions how the Higgs mechanism gives mass to the $W$ and $Z$ bosons associated with weak isospin, how photons can remain massless, and why no further massless particles arise from symmetry breaking. However, they also mention that a gauge invariant formulation of mass terms for leptons and quarks requires interaction with the Higgs field, too. Peskin and Schroeder state explicitly (see p.\,704 of \cite{PeskinSchroeder}): ``The solution to this problem [of introducing quark and lepton mass terms] will reinforce the idea that the [massless] left- and right-handed fermion fields are fundamentally independent entities, mixed to form massive fermions by some subsidiary process.'' Weinberg (see p.\,308 of \cite{WeinbergQFT2}), who was notoriously hostile to philosophy, simply states that he wants the mechanism of symmetry breaking to give masses not only to the $W$ and $Z$ bosons, but to the electron as well.

The identification of elementary particles in an ontological sense provides a key to the interpretation of quantum field theory and its low-energy limit, which is quantum mechanics. In particular, it leads to a two-world interpretation of quantum mechanics based on the strong superselection rule resulting from the postulate that states with different content of elementary particles cannot be mixed \cite{hcox1}. Such a superselection rule removes many of the paradoxes from quantum mechanics.

%\bibliography{hcopubs}


\end{document}